\input{aipcheck}

\documentclass[
    ,final            
  ]
  {aipproc}

\layoutstyle{6x9}

\usepackage{amsmath}
\usepackage{color}

\begin{document}

\title{Cosmological models with spinor and scalar fields by Noether symmetry approach}


\author{Gilberto M Kremer}{
  address={Departamento de F\'\i sica, Universidade Federal do Paran\'a,
 Curitiba, Brazil}
}

\author{Rudinei C de Souza}{
  address={Departamento de F\'\i sica, Universidade Federal do Paran\'a,
 Curitiba, Brazil}
}

\begin{abstract}
General cosmological models with spinor and scalar fields playing the role of gravitational sources are analyzed. The Noether symmetry approach is taken as a criterion to constrain the undefined potentials and couplings of the generic actions. For all the found Noether symmetries the corresponding dynamical systems can be analytically integrated. The obtained cosmological solutions describe the early and late Universe as expected by basing on the known eras of the Universe.
\end{abstract}

\maketitle


\section{Introduction}

In the inflationary paradigm \cite{1,1.1,1.2} a scalar field called \emph{inflaton} is the responsible for a rapid
accelerated expansion of the primordial Universe. The inflation solves the problems of flatness, isotropy of the microwave radiation background and unwanted relics. But the recent observations show that the present Universe also expands acceleratedly \cite{5,6}, which is not expected by the standard cosmology. The most popular attempt of explaining this acceleration is to suppose that a scalar field is also inflating the Universe today \cite{7,8}. This strange fluid that inflates the late Universe is called \emph{dark energy}. Additionally to this issue, there is an old problem related to the galaxy rotation curves \cite{2}: the common matter cannot account for the observed dynamics of the galaxies and so another type of matter is needed \cite{10}. Its nature is yet unknown and several candidates are proposed in the literature \cite{2.1}. The referred unknown matter is called \emph{dark matter}.

The possibility of a Dirac spinor field to be the source of the accelerated periods of the Universe has been explored in the literature \cite{7.0,7.3,7.3.1,7.3.2, 7.3.3}. This kind of field also produces a standard matter behavior \cite{12.1,12.2}, which may work as a model of fermionic dark matter \cite{14.0,14}. In this scope we will analyze the general role played by spinor and scalar fields in Cosmology. For this task, it will be taken general actions without specifying the potentials and couplings of the fields \emph{a priori}, since it would be generally done in an \emph{ad hoc} way. Instead of this, the undefined functions will be constrained by the requirement that the action present a Noether symmetry. Such an approach \cite{9,9.1,7.4.1,13.1,13.2, 13.3} may work as an insight on the choice of the potentials and couplings for the models, at the same time that it is a useful tool for the integration of the field equations.

The models will be analyzed for a flat Friedmann-Robertson-Walker metric with the signature $(+, \ -, \ -, \ -)$ and the associated scalar curvature defined by $$R=6\left(\frac{\ddot a}{a}+\frac{\dot a^2}{a^2}\right),$$ following the convention of the natural units $8\pi G=c=\hbar=1$.

\section{An overview of the Noether symmetry and Dirac spinor fields}

\subsection{The Noether symmetry approach}

Consider a Lagrangian of the form $\mathcal{L}=\mathcal{L}(q_k, \dot q_k)$, which is associated with the \emph{energy function}
\begin{align}
E_{\mathcal{L}}=\sum_k\frac{\partial\mathcal{L}}{\partial\dot q_k}\dot q_k-\mathcal{L},\label{funce}
\end{align}
and the following point transformation
\begin{align}
q_k\longrightarrow Q_k=q_k+\epsilon\alpha_k(\mathbf{q}),\label{trans1}
\end{align}
where the $q_k=q_k(t)$ are the generalized coordinates, the dot represents a time derivative, the $\alpha_k$ are known functions of $n$ variables -- $n$ is the number of generalized coordinates -- and $\epsilon$ is an arbitrary infinitesimal parameter.

Any invertible transformation of the generalized coordinates like (\ref{trans1}) induces a transformation of generalized velocities, such that
\begin{equation}
\dot Q^k(\textbf{q})=\frac{\partial Q^k}{\partial q_l}\dot q_l,\label{ti}
\end{equation}
where the matrix $ J=\| \partial Q^k/\partial q_l \|$ is the Jacobian of the coordinate transformation, whose determinant is supposed to be non-null. The induced transformation (\ref{ti}) is represented by the vector field
\begin{equation}
\textbf{X}=\sum_k\left(\alpha_k\frac{\partial}{\partial q_k}+
\frac{d\alpha_k}{dt}\frac{\partial}{\partial\dot q_k}\right),\label{gs}
\end{equation}
which is defined in the space tangent to the point of the transformation and is called \emph{infinitesimal generator of symmetry}. The $\alpha^k=\alpha^k(\mathbf{q})$ are the coefficients of the generator of symmetry.

The Lagrangian $\mathcal{L}$ is invariant under the transformation (\ref{gs}) if
\begin{equation}
L_\textbf{X}\mathcal{L}=0,\label{ns}
\end{equation}
with $L_\textbf{X}$ designating the Lie derivative with respect to the vector field $\textbf{X}$.

By using the E-L equations and (\ref{gs}) one can show the identity
\begin{equation}
\frac{d}{dt}\left(\sum_k\alpha_k\frac{\partial\mathcal{L}}{\partial \dot q_k}\right)=L_\textbf{X}\mathcal{L}=0,
\end{equation}
which implies that
\begin{equation}
M_0=\sum_k\alpha_k\frac{\partial\mathcal{L}}{\partial \dot q_k}, \label{constm}
\end{equation}
where $M_0$ stands for a conserved quantity (or constant of motion) associated with the transformation $\mathbf{X}$. Then it follows that if the condition (\ref{ns}) is satisfied, the Lagrangian $\mathcal{L}$ presents a Noether symmetry \cite{15.2, 15.3}.

It can be shown that the new set of variables $\left\{Q_k(\mathbf{q})\right\}$ obeys the following system of differential equations
\begin{align}
&\sum_k\left(\alpha_k\frac{\partial u_{k'}}{\partial q_k}\right)=0,\nonumber\\
&\sum_k\left(\alpha_k\frac{\partial z}{\partial q_k}\right)=1, \label{cyclicv}
\end{align}
where $k'=1, 2, ... \ k-1$ and the $u_{k'}$ and $z$ are the new coordinates $Q_k(\mathbf{q})$, with $z$ being a cyclic variable.

The Noether symmetry approach consists in taking a given Lagrangian of the form $\mathcal{L}=\mathcal{L}(q_k, \dot q_k)$ with undefined functions (generic potentials and/or couplings) and selecting their possible forms by the requirement that $\mathcal{L}$ satisfy the condition (\ref{ns}), which is performed if there is a set of $\alpha_k$ such that at least one of them is different from zero. So, if the functions that satisfy such a condition exist and are taken for the Lagrangian, it presents a Noether symmetry and there exists a conserved quantity -- given by (\ref{constm}) -- associated with its dynamics and a cyclic variable is allowed -- through equations (\ref{cyclicv}) --, which can help integrate the dynamical system.

\subsection{Dirac spinor fields in curved space-times}

The Lagrangian that describes a spinor field in a flat space-time -- Minkowski space-time --, called \emph{Dirac Lagrangian}, is the following
\begin{equation}
\mathcal{L}=\frac{\imath}{2}\ [\ \overline\psi \gamma^a \partial_a \psi-( \partial_a \overline\psi) \gamma^a
\psi \ ] - m\overline\psi\psi - V,\label{la1}
\end{equation}
where $\psi$ and $\overline\psi=\psi^{\dag}\gamma^0$ denote the spinor field and its adjoint, respectively -- the
\emph{dagger} stands for the complex conjugate transpose --, $m$ is the mass parameter and $V$ is the self-interaction potential. When a coordinate transformation is performed, the referred spinor field transforms according to the \emph{Lorentz group} \cite{17} in the following way
\begin{equation}
\psi'(x')=\exp\Big[\frac{1}{2}\lambda_{ab}\Sigma^{ab}\Big]\psi(x),
\end{equation}
where the $\lambda_{ab}$ represent the \emph{Lorentz transformation parameters} and the $\Sigma^{ab}$ are the \emph{generators} of the spinor representation, which are defined by
\begin{equation}
\Sigma^{ab}=\frac{1}{4}[\gamma^a, \gamma^b].
\end{equation}
In this definition, the $\gamma^a$ are matrixes 4x4 that satisfy the \emph{Clifford algebra}
\begin{equation}
\{\gamma^a, \gamma^b\}\equiv\gamma^a\gamma^b+\gamma^b\gamma^a=2\eta^{ab}.
\end{equation}

The generalization of the Lagrangian (\ref{la1}) for a spinor field in a curved space-time \cite{66.20,66.21a} reads
\begin{equation}
\mathcal{L}=\frac{\imath}{2}\ [\ \overline\psi \Gamma^\mu D_\mu \psi-(\overline D_\mu \overline\psi) \Gamma^\mu
\psi \ ] - m\overline\psi\psi -  V,\label{la2}
\end{equation}
being the covariant derivatives defined by
\begin{equation}
D_\mu\psi=(\partial_\mu-\Omega_\mu)\psi,
\end{equation}
\begin{equation}
\overline D_\mu\overline\psi=\partial_\mu\overline\psi+\overline\psi\Omega_\mu,
\end{equation}
where $\Omega_\mu$ is the spin connection
\begin{equation}
\Omega_\mu=-\frac{1}{4}g_{\sigma\nu}\left[\ \Gamma^\nu_{\mu\lambda}-e^\nu_b(\partial_\mu e^b_\lambda)\right ]\Gamma^\sigma\Gamma^\lambda.
\end{equation}

The generalized $\gamma$ matrixes are
\begin{equation}
\Gamma^\mu=e^\mu_a\gamma^a,
\end{equation}
which satisfy the generalized Clifford algebra
\begin{equation}
\{\Gamma^\mu, \Gamma^\nu\}=2g^{\mu\nu},
\end{equation}
with $e^a_\mu$ being called \emph{tetrad} or \emph{vierbein} and defined in the form
\begin{equation}
g_{\mu\nu}e^\mu_ae^\nu_b=\eta_{ab}, \qquad \textrm{where} \qquad e^a_\mu=\frac{\partial\xi^a}{\partial
x^\mu}.\label{tetr}
\end{equation}
The tetrad is the object that carries the information of the generalized coordinate transformation to which the spinor field is subjected.

According to the \emph{Pauli-Fierz theorem} \cite{66.22}, the potential $V$ contained in (\ref{la2}) is an exclusive function of the scalar bilinear $\Psi=\overline\psi\psi$ and the pseudo-scalar $\overline\psi\gamma^5\psi$. By simplicity, in our models we will suppose that $V$ is a function only of $\Psi$.

The E-L equations for $\overline\psi$ and $\psi$ applied to (\ref{la2}) generate
\begin{align}
&\imath\Gamma^\mu D_\mu\psi-m\psi-\frac{dV}{d\overline\psi}=0,\\
&\overline D_\mu \overline\psi\Gamma^\mu+m\overline\psi-\frac{dV}{d\psi}=0,
\end{align}
and so we have the Dirac equations for the spinor field and its adjoint in a curved space-time, i.e., in the presence of a gravitational field.

\section{Spinor field non-minimally coupled with gravity}

\subsection{Point-like Lagrangian and field equations}

The general action of a model for a Dirac spinor field non-minimally coupled
with the gravitational field through an arbitrary function of $\Psi$ reads
 \begin{equation}
S=\int\sqrt{-g}
d^4x \left\{F(\Psi)R + \frac{\imath}{2} \left[ \overline\psi
\Gamma^\mu D_\mu \psi-(\overline D_\mu \overline\psi) \Gamma^\mu
\psi  \right] - V(\Psi)\right\}, \label{1}
 \end{equation}
where $R$ is the Ricci scalar. It will be clear later why it is not supposed a massive term in this action \emph{a priori}.

For a spatially flat F-R-W metric, one can
obtain from (\ref{1}) -- by means of an integration by parts -- the following point-like Lagrangian
 \begin{equation} \mathcal{L}=6a
 \dot{a}^2F+6a^2\dot{a}\dot{\Psi}F'+\frac{\imath}{2}a^3(
 \dot{\overline\psi}\gamma^0 \psi -
 \overline\psi\gamma^0\dot{\psi})+a^3V, \label{2}
 \end{equation}
with the prime denoting a derivative with respect to $\Psi$.

From the E-L equations for $\overline\psi$ and $\psi$
applied to (\ref{2}), it follows the Dirac equations
for the spinor field and its adjoint coupled with the
gravitational field, namely,
 \begin{align}
 &\dot{\psi}+\frac{3}{2}H\psi+\imath \gamma^0\psi V'
 -\imath6(\dot{H}+2H^2)\gamma^0\psi F' =0,\label{3}\\
 &\dot{\overline\psi}+\frac{3}{2}H\overline\psi-\imath\overline\psi\gamma^0V'
 +\imath6(\dot{H}+2H^2)\overline\psi\gamma^0F'=0,\label{4}
 \end{align}
where $H=\dot{a}/a$ denotes the Hubble parameter.

The acceleration equation follows from the E-L equations
for  $a$, yielding
 \begin{equation}
 \frac{\ddot{a}}{a}=-\frac{\rho_f+3p_f}{12F},\label{6}
 \end{equation}
where the expressions for the energy density $\rho_f$ and pressure $p_f$ are defined by
\begin{align}
&\rho_f=V-6HF'\dot{\Psi}, \label{8}\\
&p_f=[V'-6(\dot{H}+2H^2)F']\Psi-V+2(F'\ddot{\Psi}+2HF'\dot{\Psi}+F''\dot{\Psi}^2).\label{9}
\end{align}

By imposing that the energy function -- see expression (\ref{funce}) -- associated with ({\ref{2}) vanishes, we obtain the modified Friedmann equation, i.e.,
 \begin{equation}
 E_\mathcal{L}=\frac{\partial \mathcal{L}}{\partial \dot{a}}
 \dot{a}+\dot{\overline\psi} \frac{\partial
 \mathcal{L}}{\partial\dot{\overline\psi}}+\frac{\partial
 \mathcal{L}}{\partial\dot{\psi}}\dot{\psi}-\mathcal{L}=0\quad\Longrightarrow\quad H^2=\frac{\rho_f}{6F}.\label{7}
 \end{equation}

\subsection{Potentials and couplings from the Noether symmetry}

In terms of the spinor components, the Lagrangian (\ref{2}) can be
written as
 \begin{equation}
\mathcal{L}\!=6a\dot{a}^2F+6a^2\dot{a}F'\sum_{i=1}^{4}\epsilon_i\Big(\dot{\psi_i^*}\psi_i+\psi_i^*
\dot{\psi_i}\Big)
+\frac{\imath}{2}a^3\sum_{i=1}^{4}\Big(\dot{\psi_i^*}\psi_i-\psi_i^*
\dot{\psi_i}\Big)+a^3V,\label{10}
 \end{equation}
and the dynamics is now described by nine coordinates, being the configuration space of the system represented by $(a, \psi_j^*, \psi_j, \dot{a}, \dot{\psi_j^*}, \dot{\psi_j})$, with $j=1,2,3,4$.

Starting from the Noether symmetry condition, $L_{\textbf{X}}\mathcal{L}=\textbf{X}\mathcal{L}=0$, applied to (\ref{10}), with $\textbf{X}$ defined for the present dynamics as
 \begin{equation}
 \textbf{X}=C_0\frac{\partial}{\partial
 a}+\dot{C_0}\frac{\partial}{\partial
 \dot{a}}+\sum_{i=1}^{4}\Bigg(C_i\frac{\partial}{\partial
 \psi_i^*}+D_i\frac{\partial}{\partial
 \psi_i}+\dot{C_i}\frac{\partial}{\partial
 \dot{\psi_i^*}}+\dot{D_i}\frac{\partial}{\partial
 \dot{\psi_i}}\Bigg)\label{12},
 \end{equation}
we arrive at an equation that depends explicitly on $\dot{a}^2$,
$\dot{a}\dot{\psi_j^*}$, $\dot{a}\dot{\psi_j}$,
$\dot{\psi_j^*}\dot{\psi_k^*}$,
$\dot{\psi_j^*}\dot{\psi_k}$,  $\dot{\psi_j}\dot{\psi_k}$,
$\dot{a}$,  $\dot{\psi_j^*}$ and $\dot{\psi_j}$, and remaining non-derivative terms. By equating
the coefficients of the above terms to zero, one obtains the
following system of coupled differential equations
 \begin{align}
 &C_0F+2a{\partial C_0\over \partial
 a}F+a^2F'\sum_{i=1}^4\left({\partial C_i\over \partial
 a}\epsilon_i\psi_i+ {\partial D_i\over \partial
 a}\epsilon_i\psi^*_i\right)
 +aF'\sum_{i=1}^4(C_i\epsilon_i\psi_i+D_i\epsilon_i\psi^*_i)=0,\label{13}
  \end{align}
 \begin{align}
 &F'\epsilon_j\psi_j\left(2C_0+a{\partial C_0\over\partial
 a}\right)+aF''\epsilon_j\psi_j\sum_{i=1}^4(C_i\epsilon_i\psi_i+D_i\epsilon_i\psi_i^*)
 +aF'D_j\epsilon_j+2F{\partial C_0\over
 \partial \psi^*_j}+\nonumber\\
 &\ \ \ \ \ \ \ \ \ \ \ \ \ \ \ \ \ \ \ \ \ \ \ \ \ \ \ \ \ \ \ \ \ \ \ \ \ \ \ \ \ \ \ \ \ \ \ \ \ \ \ \ \ \ \ \ \ \ \ \ \ \ \ \ \ \ \ \ \ \ \ \ \
 aF'\sum_{i=1}^4\left({\partial C_i\over
 \partial \psi^*_j}\epsilon_i\psi_i+ {\partial D_i\over \partial
 \psi^*_j}\epsilon_i\psi_i^*\right)=0,\label{14}
  \end{align}
 \begin{align}
 &F'\epsilon_j\psi_j^*\left(2C_0+a{\partial C_0\over \partial
 a}\right)+aF''\epsilon_j\psi_j^*\sum_{i=1}^4(C_i\epsilon_i\psi_i+D_i\epsilon_i\psi_i^*)
 +aF'C_j\epsilon_j+2F{\partial C_0\over
 \partial \psi_j}+\nonumber\\
 &\ \ \ \ \ \ \ \ \ \ \ \ \ \ \ \ \ \ \ \ \ \ \ \ \ \ \ \ \ \ \ \ \ \ \ \ \ \ \ \ \ \ \ \ \ \ \ \ \ \ \ \ \ \ \ \ \ \ \ \ \ \ \ \ \ \ \ \ \ \ \ \ \ \
 aF'\sum_{i=1}^4\left({\partial C_i\over \partial
 \psi_j}\epsilon_i\psi_i+ {\partial D_i\over \partial
 \psi_j}\epsilon_i\psi_i^*\right)=0,\label{15}
  \end{align}
 \begin{align}
 &F'\left({\partial
 C_0\over \partial \psi_k^*}\epsilon_j\psi_j+ {\partial C_0\over
 \partial \psi_j^*}\epsilon_k\psi_k\right)=0,\qquad
 F'\left({\partial C_0\over \partial \psi_k}\epsilon_j\psi_j^*+
 {\partial C_0\over \partial \psi_j}\epsilon_k\psi_k^*\right)=0,\label{16}
  \end{align}
 \begin{align}
 &F'\left({\partial C_0\over \partial
 \psi_k}\epsilon_j\psi_j+ {\partial C_0\over \partial
 \psi_j^*}\epsilon_k\psi_k^*\right)=0,\qquad
 \sum_{i=1}^4\left({\partial C_i\over \partial a}\psi_i- {\partial
 D_i\over \partial a}\psi_i^*\right)=0,\label{17}
  \end{align}
 \begin{align}
 &3C_0\psi_j+aD_j+a\sum_{i=1}^4\left({\partial C_i\over \partial
 \psi_j^*}\psi_i- {\partial D_i\over \partial
 \psi_j^*}\psi_i^*\right)=0,\label{18}
  \end{align}
 \begin{align}
 &3C_0\psi_j^*+aC_j-a\sum_{i=1}^4\left({\partial C_i\over
 \partial \psi_j}\psi_i- {\partial D_i\over \partial
 \psi_j}\psi_i^*\right)=0,\label{19}
  \end{align}
 \begin{align}
 &3C_0V+aV'\sum_{i=1}^4\left(C_i\epsilon_i\psi_i+D_i\epsilon_i\psi_i^*\right)=0.\label{20}
 \end{align}
 In equations (\ref{13}) through (\ref{20}) it was introduced the symbol
\begin{align}\nonumber
&\epsilon_i=+1\qquad\hbox{for}\qquad i=1,2;\\
&\epsilon_i=-1\qquad\hbox{for}\qquad i=3,4.
\end{align}

The system (\ref{13})-(\ref{20}) is solved for two cases: $F'=0$ and $F'\neq0$ -- the details of the integration of this system can be found in the reference \cite{66.12}. The corresponding solutions are exposed as follows.

\paragraph{Case $F'=0$}

 \begin{align}
 &C_0={k\over a^{1/2}},\qquad C_j=-{3\over2}k{\psi_j^*\over a^{3/2}}+\beta\epsilon_j\psi_j^*,\qquad D_j=-{3\over2}k{\psi_j\over a^{3/2}}-\beta\epsilon_j\psi_j,\nonumber\\
 &F=\textrm{constant},\qquad V=\lambda\Psi,\label{23}
 \end{align}
with $k, \beta, \lambda$= constant.

\paragraph{Case $F'\neq0$}
\begin{equation}
\textrm{Set 1}\nonumber
\end{equation}
 \begin{align}
 &C_0=ka,\qquad C_j=-{3\over2}k{\psi_j^* }+\beta\epsilon_j\psi_j^*, \qquad D_j=-{3\over2}k{\psi_j }-\beta\epsilon_j\psi_j,\nonumber\\
 &F=\alpha\Psi,\qquad V=\lambda\Psi,
 \end{align}
\begin{equation}
\textrm{Set 2}\nonumber
\end{equation}
 \begin{align}
 &C_0={k\over a},\qquad C_j=-{3\over2}k{\psi_j^* }+\beta\epsilon_j\psi_j^*,\qquad D_j=-{3\over2}k{\psi_j }-\beta\epsilon_j\psi_j,\nonumber\\
 &F=\alpha\Psi^{1/3},\qquad V=\lambda\Psi,
 \end{align}
where $\alpha$ is a constant.

\subsection{Solving the field equations}

We cannot distinguish physically the potential $V$ given by
$(\ref{23})_2$ from a mass term -- see (\ref{la2}). Then one can consider $V = \lambda\Psi \equiv m\Psi$, so that $\lambda$
corresponds to a mass parameter. Thus from now on $\lambda$ will be replaced by $m$.

From (\ref{3}) and (\ref{4}) one can
build an evolution equation for $\Psi$, which reads
\begin{equation}
 \dot\Psi+3H\Psi=0,\qquad \hbox{so that}\qquad \Psi={\Psi_0\over
 a^3},\label{28a}
 \end{equation}
 where $\Psi_0$ is a constant. Then, once $\Psi=\Psi(a)$, equation (\ref{7}) becomes a function only of $a$ and, as it will be seen below, it can be directly integrated for all the cases. Therefore it is not necessary to calculate the constants of motion.

\paragraph{Case $F'=0$}

The energy density and pressure of the field follow from
(\ref{8}) and (\ref{9}) through $\ref{28a}$, yielding
 \begin{equation}
 \rho_f={m\Psi_0\over a^3},\qquad p_f=0,
 \end{equation}
which imposes that $\Psi_0>0$.

 The choice  $F=$ constant $=1/2$ corresponds to
a minimal coupling between the spinor and gravitational fields. In so doing, by inserting the obtained $\rho_f$ into (\ref{7}), it
furnishes the scale factor
 \begin{equation}
 a(t)=a_0t^{2/3}, \qquad\hbox{where}\qquad
 a_0=\left(3\over2\right)^{2/3}\left(m\Psi_0\over3\right)^{1/3},\label{solA}
 \end{equation}
 which describes a decelerated Universe as it was dominated by a standard pressureless matter field.

 Substituting (\ref{solA}) in (\ref{3}), we have the solution for the spinor field
 \begin{equation}
 \psi(t)=a_0^{-3/2}\left(
\begin{array}{c}
\psi_{1}^0e^{-\imath mt}  \\
\psi_{2}^0e^{-\imath mt}  \\
\psi_{3}^0e^{\imath mt}  \\
\psi_{4}^0e^{\imath mt}  \\
\end{array}
\right)t^{-1},\label{spi1}
\end{equation}
with the $\psi_j^0$ being constants such that $\Psi_0=\psi_1^{*0}\psi_1^0+\psi_2^{*0}\psi_2^0-\psi_3^{*0}\psi_3^0-\psi_4^{*0}\psi_4^0$.

\paragraph{Case $F'\neq0$}
For the Set 1 solutions, when $F=\alpha\Psi$, the energy density and pressure read
 \begin{equation}
 \rho_f=-{m\Psi_0\over
 2a^3},\qquad p_f=-\rho_f,\label{33a}
 \end{equation}
 so the integration of (\ref{7}) gives
 \begin{equation}
 a(t)=e^{Kt},\qquad\hbox{where}\qquad K=\sqrt{-{{m\over12\alpha}}},\label{dss}
 \end{equation}
which trough (\ref{3}) renders the following solution
 \begin{equation}
 \psi(t)=\left(
\begin{array}{c}
\psi_{1}^0e^{-2\imath m t}  \\
\psi_{2}^0e^{-2\imath m t}  \\
\psi_{3}^0e^{2\imath m t}  \\
\psi_{4}^0e^{2\imath m t}  \\
\end{array}
\right)e^{-\frac{3K}{2}t},\label{spi2}
\end{equation}
with the $\psi_j^0$ following the same constraint as for solution (\ref{spi1}).

We infer from (\ref{33a})$_1$ that one must have $\Psi_0<0$, which imposes that $\alpha<0$ since the coupling
$F=\alpha\Psi$ is a positive quantity. And by its turn, the condition $\alpha<0$ implies that $K$ ia a real quantity.
So the solution (\ref{dss}) can describe a De Sitter-like Universe. Hence in this case we could identify the spinor field with the "inflaton". The result (\ref{33a})$_2$ asserts that the pressure is always the negative of its corresponding energy density, similarly to the cosmological constant state equation.

For the Set 2 solutions, when $F=\alpha\Psi^{1/3}$, equation (\ref{7}) does not have a solution.

\section{Scalar and spinor fields minimally coupled with gravity}

\subsection{General action and field equations}

Consider a general action for a scalar field and a spinor field minimally coupled with the gravity

\begin{align}\nonumber
S=\int d^4x\sqrt{-g}\ \Bigg\{\frac{R}{2}&+\frac{1}{2}g^{\mu\nu}\partial_\mu\phi\partial_\nu\phi-U(\phi)+\frac{\imath}{2}\left[\overline\psi\Gamma^\mu D_\mu\psi-(\overline D_\mu\overline\psi)\Gamma^\mu\psi\right]-V(\Psi)\Bigg\}\\
&+\int d^4x\sqrt{-g}\mathcal{L}_M,\nonumber\\
\label{ga}
\end{align}
where $\mathcal{L}_M$ stands for the Lagrangian of a common matter field.

For a flat F-R-W metric applied to (\ref{ga}), with the fields $\phi$ and $\psi$ being spatially homogeneous and the common matter pressureless, we can write the point-like Lagrangian
\begin{equation}
\mathcal{L}=3a\dot{a}^2-a^3\left(\frac{\dot\phi^2}{2}-U\right)+\frac{\imath}{2}a^3\left(
\dot{\overline\psi}\gamma^0 \psi -
\overline\psi\gamma^0\dot{\psi}\right)+a^3V+\rho_M^0, \label{plL}
\end{equation}
where $\rho_M^0$ denotes the energy density of the common matter at an initial instant.

Imposing that the energy function associated with (\ref{plL}) is null,
the result is the Friedmann equation
\begin{align}\nonumber
&\ E_\mathcal{L}=\frac{\partial \mathcal{L}}{\partial \dot{a}}
\dot{a}+\frac{\partial \mathcal{L}}{\partial \dot{\phi}}
\dot{\phi}+\dot{\overline\psi} \frac{\partial
\mathcal{L}}{\partial\dot{\overline\psi}}+\frac{\partial
\mathcal{L}}{\partial\dot{\psi}}\dot{\psi}-\mathcal{L}=0
\\ \Longrightarrow\ &\ H^2=\frac{\rho_M+\rho_\psi+\rho_\phi}{3},\label{fride}
\end{align}
with $\rho_\psi$ and $\rho_\phi$ denoting the energy densities, which are defined as
\begin{equation}
\rho_\psi=V, \qquad \rho_\phi=\frac{1}{2}\dot\phi^2+U. \label{ed}
\end{equation}

From the E-L equations for $a$ applied to (\ref{plL}), we have the acceleration equation
\begin{equation}
\frac{\ddot{a}}{a}=-\frac{\rho_M+\rho_\psi+\rho_\phi+3\left(p_\psi+p_\phi\right)}{6}, \label{acce}
\end{equation}
and for $\overline\psi$ and $\psi$, the Dirac equations
for the spinor field and its adjoint coupled with the gravitational field, respectively,
\begin{align}
&\dot{\psi}+\frac{3}{2}H\psi+\imath\gamma^0\psi\frac{dV}{d\Psi}=0,\\
&\dot{\overline\psi}+\frac{3}{2}H\overline\psi-\imath\overline\psi\gamma^0\frac{dV}{d\Psi}=0.
\end{align}
In (\ref{acce}) the pressures are defined as
\begin{equation}
p_\psi=\Psi\frac{dV}{d\Psi}-V, \qquad p_\phi=\frac{\dot\phi^2}{2}-U. \label{press}
\end{equation}

The E-L equations for $\phi$ furnishes
\begin{equation}
\ddot\phi+3H\dot\phi+\frac{dU}{d\phi}=0,\label{KGe}
\end{equation}
and so we have the Klein-Gordon equation for the field $\phi$.

\subsection{Determination of the Noether potentials}

Expressing the Lagrangian (\ref{plL}) in terms of the spinor components, it takes the form
\begin{equation}
\mathcal{L}=3a\dot{a}^2-a^3\left(\frac{\dot\phi^2}{2}-U\right)+\frac{\imath}{2}a^3\sum_{i=1}^{4}\left(\dot{\psi_i^*}\psi_i-\psi_i^*
\dot{\psi_i}\right)+a^3V+\rho_M^0. \label{plL1}
\end{equation}
Thus the configuration space of the system is represented by $(a, \phi, \psi_j^*, \psi_j, \dot{a}, \dot\phi, \dot{\psi_j^*}, \dot{\psi_j})$.

The Noether symmetry condition,
$L_{\textbf{X}}\mathcal{L}=\textbf{X}\mathcal{L}=0$, applied to (\ref{plL1}), with $\textbf{X}$ defined for our problem as
\begin{eqnarray}
\textbf{X}=C_0\frac{\partial}{\partial
a}+\dot{C_0}\frac{\partial}{\partial
\dot{a}}+D_0\frac{\partial}{\partial
\phi}+\dot{D_0}\frac{\partial}{\partial
\dot{\phi}}+\sum_{i=1}^{4}\left(C_i\frac{\partial}{\partial
\psi_i^*}+D_i\frac{\partial}{\partial
\psi_i}+\dot{C_i}\frac{\partial}{\partial
\dot{\psi_i^*}}+\dot{D_i}\frac{\partial}{\partial
\dot{\psi_i}}\right),
\end{eqnarray}
renders the following system of coupled differential equations
\begin{equation}
C_0+2a\frac{\partial C_0}{\partial a}=0, \qquad 3C_0+2a\frac{\partial D_0}{\partial \phi}=0, \qquad 6\frac{\partial C_0}{\partial \phi}-a^2\frac{\partial D_0}{\partial a}=0, \label{sc1}
\end{equation}
\begin{equation}
\sum_{i=1}^4\left({\partial C_i\over \partial a}\psi_i- {\partial
D_i\over \partial a}\psi_i^*\right)=0, \qquad \sum_{i=1}^4\left({\partial C_i\over \partial \phi}\psi_i- {\partial
D_i\over \partial \phi}\psi_i^*\right)=0, \label{sc2}
\end{equation}
\begin{equation}
3C_0\psi_j+aD_j+a\sum_{i=1}^4\left({\partial C_i\over \partial
\psi_j^*}\psi_i- {\partial D_i\over \partial
\psi_j^*}\psi_i^*\right)=0, \label{sc3}
\end{equation}
\begin{equation}
3C_0\psi_j^*+aC_j-a\sum_{i=1}^4\left({\partial C_i\over
\partial \psi_j}\psi_i- {\partial D_i\over \partial
\psi_j}\psi_i^*\right)=0, \label{sc4}
\end{equation}
\begin{equation}
\frac{\partial C_0}{\partial \psi_j^*}=0, \qquad \frac{\partial C_0}{\partial \psi_j}=0,\qquad \frac{\partial D_0}{\partial \psi_j^*}=0, \qquad \frac{\partial D_0}{\partial \psi_j}=0, \label{sc5}
\end{equation}
\begin{equation}
3C_0(U+V)+aD_0\frac{dU}{d\phi}+a\sum_{i=1}^4\left(C_i\epsilon_i\psi_i+D_i\epsilon_i\psi_i^*\right)\frac{dV}{d\Psi}=0. \label{sc6}
\end{equation}

 One can see from equations (\ref{sc5}) that the coefficients $C_0$ and $D_0$ are functions only of $a$ and $\phi$. Then, assuming that $C_0$ and $D_0$ are separable functions
\begin{equation}
C_0=c_1(a)c_2(\phi), \qquad D_0=d_1(a)d_2(\phi),
\end{equation}
we obtain the solution for the system (\ref{sc1})-(\ref{sc6})
\begin{equation}
C_0=\frac{Ae^{\alpha\phi}+Be^{-\alpha\phi}}{\sqrt{6a}},
\end{equation}
\begin{equation}
D_0=-\frac{Ae^{\alpha\phi}-Be^{-\alpha\phi}}{a^{3/2}},
\end{equation}
\begin{equation}
C_j=-\alpha\frac{Ae^{\alpha\phi}+Be^{-\alpha\phi}}{a^{3/2}}{\psi_j^* }+\beta\epsilon_j\psi_j^*,
\end{equation}
\begin{equation}
D_j=-\alpha\frac{Ae^{\alpha\phi}+Be^{-\alpha\phi}}{a^{3/2}}{\psi_j}-\beta\epsilon_j\psi_j,
\end{equation}
\begin{equation}
U=U_0\left(Ae^{\alpha\phi}-Be^{-\alpha\phi}\right)^2, \qquad V=V_0\Psi, \label{potentials}
\end{equation}
with $A, B, U_0, V_0$ and $\beta$ being constants and $\alpha=\frac{1}{2}\sqrt{\frac{3}{2}}$.

\paragraph{Dark matter as femionic particles}

The Noether potential $V=V_0\Psi$ is essentially a term of mass. Then, as before, we replace the constant $V_0$ by $m$. And from $(\ref{ed})_1$ and $(\ref{press})_1$ the energy density and pressure are
\begin{equation}
\rho_\psi=m\Psi, \qquad p_\psi=0,\label{dmatt}
\end{equation}
characterizing a field of pressureless matter -- when $\Psi$ is a non-negative function.

This result \emph{suggests} that the spinor field behaves as a standard matter field. But obviously this field has a nature different from that of the common matter since it describes a "fluid" exclusively composed by fermionic particles. Such a field produces an additional pressureless matter that may be identified with the dark matter. For this propose, we must assume that these fermionic particles interact only gravitationally with the common matter or have very weak non-gravitational interactions with it. On the other hand, the potential $(\ref{potentials})_1$ can produce an accelerated expansion. Therefore we may identify the field $\phi$ with the dark energy. Hence, from this point on, the dark sector will be identified with the fields $\phi$ and $\psi$.

\subsection{Analytical solutions of the field equations}

Taking $U$ and $V$ given by (\ref{potentials}) for (\ref{fride})-(\ref{KGe}), we have the following system to solve
\begin{equation}
3H^2=\frac{\rho_M^0}{a^3}+m\Psi+\frac{\dot\phi^2}{2}+U_0\left(Ae^{\alpha\phi}-Be^{-\alpha\phi}\right)^2, \label{fe}
\end{equation}
\begin{equation}
H^2+2\frac{\ddot{a}}{a}+\frac{\dot\phi^2}{2}-U_0\left(Ae^{\alpha\phi}-Be^{-\alpha\phi}\right)^2=0, \label{ae}
\end{equation}
\begin{equation}
\ddot\phi+3H\dot\phi+2\alpha U_0\left(A^2e^{2\alpha\phi}-B^2e^{-2\alpha\phi}\right)=0, \label{kge}
\end{equation}
\begin{equation}
\dot{\psi}+\frac{3}{2}H\psi+\imath m\gamma^0\psi=0, \qquad \dot{\overline\psi}+\frac{3}{2}H\overline\psi-\imath m\overline\psi\gamma^0=0. \label{dirac1,2}
\end{equation}

  Once the dynamical system now presents a Noether symmetry, one has an additional equation provided by the constant of motion, namely
\begin{align}\nonumber
M_0&=C_0\frac{\partial\mathcal{L}}{\partial
\dot{a}}+D_0\frac{\partial\mathcal{L}}{\partial
\dot{\phi}}+\sum_{i=1}^{4}\left(C_i\frac{\partial\mathcal{L}}{\partial
\dot{\psi_i^*}}+D_i\frac{\partial\mathcal{L}}{\partial
\dot{\psi_i}}\right)\nonumber\\
&=\left(Ae^{\alpha\phi}+Be^{-\alpha\phi}\right)\sqrt{6a}\dot a+\left(Ae^{\alpha\phi}-Be^{-\alpha\phi}\right)a^{3/2}\dot\phi+\imath\beta\ a^3\Psi, \label{cm}
\end{align}
which is determined from (\ref{constm}). In this equation one can make $\beta=\imath\beta_0$, with $\beta_0$ being a real constant, in order to have a constant of motion whose value is real.

 Consider now a transformation of variables that changes the set of coordinates $\{a, \phi, \psi_j^*, \psi_j\}$ to $\{z, u, v_j, w_j\}$, which satisfies the system
\begin{equation}
C_0\frac{\partial u}{\partial
a}+D_0\frac{\partial u}{\partial
\phi}+\sum_{i=1}^{4}\left(C_i\frac{\partial u}{\partial
\dot{\psi_i^*}}+D_i\frac{\partial u}{\partial \dot{\psi_i}}\right)=0,\label{v1}
\end{equation}
\begin{equation}
C_0\frac{\partial v_j}{\partial
a}+D_0\frac{\partial v_j}{\partial
\phi}+\sum_{i=1}^{4}\left(C_i\frac{\partial v_j}{\partial
\dot{\psi_i^*}}+D_i\frac{\partial v_j}{\partial
\dot{\psi_i}}\right)=0,\label{v2}
\end{equation}
\begin{equation}
C_0\frac{\partial w_j}{\partial
a}+D_0\frac{\partial w_j}{\partial
\phi}+\sum_{i=1}^{4}\left(C_i\frac{\partial w_j}{\partial
\dot{\psi_i^*}}+D_i\frac{\partial w_j}{\partial
\dot{\psi_i}}\right)=0,\label{v3}
\end{equation}
\begin{equation}
C_0\frac{\partial z}{\partial
a}+D_0\frac{\partial z}{\partial
\phi}+\sum_{i=1}^{4}\left(C_i\frac{\partial z}{\partial
\dot{\psi_i^*}}+D_i\frac{\partial z}{\partial
\dot{\psi_i}}\right)=1,\label{v4}
\end{equation}
with $z$ being the cyclic variable. This system is obtained by applying (\ref{cyclicv}).

Using the constant of motion (\ref{cm}) and a transformation of variables satisfying (\ref{v1})-(\ref{v4}), we will look for a solution to the field equations (\ref{fe})-(\ref{dirac1,2}).

Firstly, equations $(\ref{dirac1,2})_1$ and $(\ref{dirac1,2})_2$ can be reduced to an equation for $\Psi$ with the same form of (\ref{28a}), then the form of the bilinear is again $\Psi=\Psi_0/a^3$, so that $\Psi_0>0$ by (\ref{dmatt}). And by solving $(\ref{dirac1,2})_1$ in terms of $a$, the solution for $\psi$ reads
\begin{equation}
 \psi=\left(%
\begin{array}{c}
\psi_{1}^0e^{-\imath mt}  \\
\psi_{2}^0e^{-\imath mt}  \\
\psi_{3}^0e^{\imath mt}  \\
\psi_{4}^0e^{\imath mt}  \\
\end{array}%
\right)a^{-3/2}\label{spinor1},
\end{equation}
with the $\psi_j^0$ subjected to the same constraint of (\ref{spi1}). The solution $\Psi=\Psi_0/a^3$ through (\ref{dmatt}) implies the energy density $\rho_\psi=m\Psi_0/a^3$, so the extra matter field is naturally added with the common matter field, which corroborates the possible identification of the spinor field with the dark matter.

Since equations (\ref{dirac1,2}) can be independently solved in terms of $a$, we effectively have a reduction of the system (\ref{fe})-(\ref{dirac1,2}) to equations that involve only the variables $a$ and $\phi$. Thus we will solve the problem through the reduced system
\begin{equation}
3H^2=\frac{\rho_M^0+m\Psi_0}{a^3}+\frac{\dot\phi^2}{2}+U_0\left(Ae^{\alpha\phi}-Be^{-\alpha\phi}\right)^2,
\end{equation}
\begin{equation}
\ddot\phi+3H\dot\phi+2\alpha U_0\left(A^2e^{2\alpha\phi}-B^2e^{-2\alpha\phi}\right)=0,
\end{equation}
\begin{equation}
\left(Ae^{\alpha\phi}+Be^{-\alpha\phi}\right)\sqrt{6a}\dot a+\left(Ae^{\alpha\phi}-Be^{-\alpha\phi}\right)a^{3/2}\dot\phi=\overline{M}_0,
\end{equation}
where $\overline{M}_0=M_0+\beta_0\Psi_0$ and the result $\Psi=\Psi_0/a^3$ was used.

Hence we need to find a new set of variables related only to $a$ and $\phi$. An adequate particular solution of this type for the system (\ref{v1})-(\ref{v4}) is
\begin{equation}
u=a^{3/2}\frac{Ae^{\alpha\phi}-Be^{-\alpha\phi}}{\sqrt{6}AB}, \qquad v_j=w_j=0, \qquad z=a^{3/2}\frac{Ae^{\alpha\phi}+Be^{-\alpha\phi}}{\sqrt{6}AB}. \label{change2}
\end{equation}

Using (\ref{change2}) we can express the constant of motion in terms of the cyclic variable
\begin{equation}
\dot z = \frac{\overline{M}_0}{4AB}, \qquad \textrm{so that} \qquad z(t) = z_1t+z_2, \label{sol1}
\end{equation}
with $z_1=\overline{M}_0/4AB$ and $z_2=$ constant.

Equation (\ref{fe}) takes the following form in the new variables
\begin{equation}
\dot z^2 = \dot u^2 + 3ABU_0u^2 + \frac{\rho_0+m\Psi_0}{2AB}.
\end{equation}
Substituting $\dot z$ from (\ref{sol1}) in the above equation, we get its integration and obtain
\begin{equation}
u(t)=u_0\sin{(\omega t+b_0)},
\end{equation}
where
\begin{equation}
u_0^2=\frac{\overline{M}_0^2-8AB\left(\rho_M^0+m\Psi_0\right)}{48A^3B^3U_0}, \qquad \omega^2=3ABU_0,\label{constants}
\end{equation}
with $b_0$ being a constant.

From the solutions $z(t)$ and $u(t)$, expressed in terms of the original variables through the relations (\ref{change2}), we get the explicit forms of $a(t)$ and $\phi(t)$
\begin{equation}
a(t)=\left(\frac{\omega^2}{2U_0}\right)^{1/3}\left\{z_1^2t^2+2z_1z_2t+z_2^2-u_0^2\sin^2{\left(\omega t+b_0\right)}\right\}^{1/3}, \label{solution1}
\end{equation}
\begin{equation}
\phi(t)=\frac{1}{2\alpha}\ln{\left\{\frac{z_1t+z_2+u_0\sin{\left(\omega t+b_0\right)}}{z_1t+z_2-u_0\sin{\left(\omega t+b_0\right)}}\right\}}-\frac{1}{2\alpha}\ln{\left(\frac{A}{B}\right)}.\label{solution2}
\end{equation}

Now from (\ref{spinor1}) one can write the time evolution of the spinor field
\begin{equation}
 \psi(t)=\frac{\sqrt{2U_0}}{\omega}\left(%
\begin{array}{c}
\psi_{1}^0e^{-\imath mt}  \\
\psi_{2}^0e^{-\imath mt}  \\
\psi_{3}^0e^{\imath mt}  \\
\psi_{4}^0e^{\imath mt}  \\
\end{array}
\right)\Theta(t),\label{solution3}
\end{equation}
where
\begin{equation}
\Theta(t)=\left\{z_1^2t^2+2z_1z_2t+z_2^2-u_0^2\sin^2{\left(\omega t+b_0\right)}\right\}^{-1/2}.
\end{equation}

The solution (\ref{solution1}) describes a Universe with an oscillatory expansion rate and can account for the transition from a decelerated to an accelerated era, in according to the observations. Further, it delineates a Universe that will return to a decelerated period and turn to accelerate in the future and so on, such that these alternated periods of acceleration and deceleration have a behavior like that from a deadened oscillator -- when $t\rightarrow\infty$, one has $\ddot a\rightarrow0$. The time evolution of the scalar field (\ref{solution2}) makes the pressure $(\ref{press})_2$ oscillate between a negative and a positive value, and so the field alternately behaves as dark energy and matter, being the cause of the oscillatory expansion. The detailed analysis of the above solutions can be found in the reference \cite{66.13}.

\section{Conclusions}

Generic cosmological models with spinor and scalar fields were constrained through the Noether symmetry method. The proprieties of the Noether symmetry rendered the complete integration of the dynamical systems. The cosmological solutions showed that the spinor field can play the role of "inflaton" -- non-minimally coupled case --, such that its equation of state is similar to that of the cosmological constant; and for the minimally coupled case, a behavior of standard matter was produced. For the model with spinor and scalar fields minimally coupled, the spinor field presented a standard matter behavior -- which may describe the dark matter -- and the scalar field an oscillating equation of state, which passed from a matter-like to a dark energy-like description producing alternated periods of accelerated and decelerated expansion.


\begin{theacknowledgments}
The authors acknowledge the support of Capes and Cnpq, Brazil.
\end{theacknowledgments}



\bibliographystyle{aipproc}   


\IfFileExists{\jobname.bbl}{}
 {\typeout{}
  \typeout{******************************************}
  \typeout{** Please run "bibtex \jobname" to optain}
  \typeout{** the bibliography and then re-run LaTeX}
  \typeout{** twice to fix the references!}
  \typeout{******************************************}
  \typeout{}
 }


\end{document}